\documentclass[12pt]{article}
\usepackage{amsmath,amssymb,amsfonts,bm}
\usepackage{fullpage,xcolor,url,enumitem,natbib}

\newtheorem{lemma}{Lemma}
\newtheorem{proposition}{Proposition}

\def\pconv{\smash{\mathop{\longrightarrow}\limits^p}}     
\def\dconv{\smash{\mathop{\longrightarrow}\limits^d}}     

\newcommand{\lio}{\Lambda_i^0}
\newcommand{\fto}{F_t^0}
\newcommand{\pr}{^\prime}

\newcommand{\sumiN}{\sum_{i=1}^N}
\newcommand{\sumtT}{\sum_{t=1}^T}
\newcommand{\sumsT}{\sum_{s=1}^T}

\newcommand{\OpdNTitwo}{O_p(\delta_{NT}^{-2})}
\newcommand{\OpdNTione}{O_p(\delta_{NT}^{-1})}

\newcommand{\Fo}{\bm F^0}
\newcommand{\Fop}{\bm F^{0'}}
\newcommand{\Lo}{\bm \Lambda^0}
\newcommand{\Lop}{\bm \Lambda^{0'}}

\newcommand{\Hpi}{\bm H_{NT}^{'^{-1}}}
\newcommand{\norm}[1]{\left\Vert#1\right\Vert}
\def\Lcpca{\bar{ \bm \Lambda}_{\gamma,\infty}}
\def\Lridge{\bar{\bm \Lambda}_{\gamma,0}}

\def\Fpca{\hat{\bm F}}
\def\Lpca{\hat{\bm \Lambda}}
\def\Dpca{\bm D_{NT,r}}
\def\Frpca{\bar{\bm F}}
\def\Lrpca{\bar{\bm \Lambda}}
\def\argmin{\text{argmin}}
\def\Lp{\bm \Lambda^{0\pr}}
\def\Fp{\bm F^{0\pr}}
\def\vec{\mathrm{vec}}

\def\Lapca{\tilde{\bm \Lambda}}
\def\Lpca{\hat{\bm \Lambda}}

\newcommand{\tF}{\tilde {\bm F}}

\newcommand{\tL}{\tilde {\bm \Lambda}}
\newcommand{\tFp}{\tilde {\bm F}^\prime}
\newcommand{\tLp}{\tilde {\bm \Lambda}^\prime}

\begin{document}
\title{\large{\textsc{Simpler Proofs For Approximate Factor Models of Large Dimensions}}}
\author{Jushan Bai\thanks{Columbia University, 420 W. 118 St. MC 3308, New York, NY 10027.
Email: jb3064@columbia.edu}
  \and Serena Ng\thanks{Columbia University and NBER, 420 W. 118 St. MC 3308,
  New York, NY 10027. Email: serena.ng@columbia.edu.
 This work is supported by the National Science
Foundation  SES-2018369 (Ng).}}
\date{\today\bigskip}

\maketitle
\begin{abstract}
   Estimates of the approximate factor model are increasingly used in empirical work. Their theoretical properties, studied some twenty years ago, also laid the ground work for analysis on large dimensional panel data models with cross-section dependence. This paper presents simplified proofs for the estimates  by using alternative rotation matrices, exploiting properties of low rank matrices, as well as the singular value decomposition of the data in addition to its covariance structure.  These simplifications facilitate interpretation of results and provide a more friendly introduction to researchers new to the field.    New results are provided to  allow  linear restrictions to be imposed on  factor models.
\end{abstract}



\thispagestyle{empty}
\setcounter{page}{0}
\newpage
\baselineskip=18.0pt

\section{Introduction}

An active area of  research in the last twenty years is analysis of  panel  data  with cross-section dependence, where the panel has dimension $T\times N$, and where  $T$ (the time) and $N$ (the cross-section) dimensions are both large.  Classical factor models studied by \citet{anderson-rubin} and \citet{lawley-maxwell-74} among others are designed to capture cross-section dependence when either $T$ or $N$ is fixed, and that  errors are iid across time and units. The  {\it approximate factor model} formulated in \citet{chamberlain-rothschild}  relaxes many these assumptions, so what remains is to be to able  take the theory to the data. \cite{connor-korajczyk-93}  suggest to estimate the factors by the method of asymptotic principal components (APC). Consistency proofs were subsequently given in  \cite{stock-watson-jasa:02}, \citet{baing-ecta:02} under the  assumption that  $N,T
\rightarrow\infty$ with $\sqrt{N}/T\rightarrow \infty$. \citet{baing-ecta:06} provide the conditions under which the factor estimates can be treated in subsequent regressions as though they were observed. Novel uses of the factor estimates such as diffusion index forecasting pioneered in  \citet{stock-watson-diforc}) and  factor-augmented autoregressions such as considered in \citet{bernanke-boivin-eliasz}, along with the natural role  that common factors play  in many theoretical models in economics and finance have contributed to the popularity of large dimensional factor analysis.

 Arguably, the three fundamental results in this literature are i) the  consistency proof of  the estimated factor space at rate $\min(N,T)$,  ii) consistent estimation of the number of factors,  and  (iii)   $\sqrt{N}, \sqrt{T},$ and  $\min(\sqrt{N},\sqrt{T})$ asymptotic normality of  the estimated factors, the loadings, and the common component, respectively.   The point of departure in these results, given \citet{baing-ecta:02} and \citet{bai-ecta-03},  is an analysis of  the factor estimates relative to a specific rotation of the true factors first considered in \citet{stock-watson-di-wp} that is defined from the covariance structure of the data.  This leads to a decomposition of the estimation error  into  four terms  and carefully deriving  the limit for  each of them. Though a large body of research is built on  these theoretical results, the arguments are lengthy and often not particularly intuitive. 

In this paper, we show that the key  results can be obtained using simpler arguments and under higher level assumptions. It turns out  that   inspection of the norm of the $T\times T$ population covariance of the errors  is already sufficient to establish that the factor space can be consistently estimated at rate  $\min(N,T)$ from which consistent estimation of the number of factors can be easily established.  Exploiting the eigen-decomposition of the data and not only its covariance leads to different representation of the factor estimates that also simplify the analysis. Most important is the  recognition that the rotation matrix is not unique. We present four asymptotically equivalent rotation matrices that  simplify the proofs for asymptotic normality. It will be shown that the asymptotic variance of the factor estimates can be represented in many ways. This  little known fact makes it possible to conduct inference using an estimate of the variance that the researcher finds most computationally  convenient.  The simplified arguments, presented in consistent notation,  should help students and researchers new to the field better understand  the role that large $N$ and $T$ play in estimation of  approximate factor models.

Economic analysis sometimes impose specific restrictions on the model.  Because we can only estimate the factor space up to a rotation matrix, the problem is a bit more tricky.  We provide results for estimation of  factor models with linear restrictions These results should be of interest as factor estimation finds more ways into  economic applications.

 \section{Model Setup and Assumptions}

  We use  $i=1,\ldots N$ to index cross-section units and $t=1,\ldots T$ to index time series observations. Let  $X_i=(X_{i1},\ldots X_{iT})^\prime$ be a $T\times 1$ vector of random variables and
 $\bm X=(X_1,X_2,\ldots, X_N)$ be a $T\times N$  matrix.  In practice, $X_{i}$ is transformed to be stationary, demeaned, and often standardized. The normalized data $\bm Z=\frac{\bm X}{\sqrt{NT}}$
has  singular value decomposition (\textsc{svd})
\[ \bm Z=\frac{\bm X}{\sqrt{NT}}
 =\bm U_{NT}\bm  D_{NT}\bm V_{NT}\pr=\bm U_{NT,r} \bm D_{NT,r} \bm V_{NT,r}\pr+ \bm U_{NT,N-r} \bm  D_{NT,N-r} \bm V_{NT,N-r}\pr.\]
In the above, $\bm  D_{NT,r}$ is a diagonal matrix of $r$ singular values $d_{NT,1},\ldots, d_{NT,r}$ arranged in descending order, $\bm U_{NT,r}, \bm V_{NT,r}$ are the corresponding left and right singular vectors respectively.  Note that while the $r$ large singular values of $\bm X$  diverge and the remaining $N-r$ ones are bounded, the $r$ largest singular values
 of $\bm Z$ are bounded and the remaining ones tend to zero because the singular values of $\bm Z$ are those of $\bm X$ divided by $\sqrt{NT}$. The   \citet{eckart-young} theorem    posits that  the best  rank $k$ approximation of $\bm Z$  is $\bm U_{NT,k}\bm D_{NT,k}\bm V_{NT,k}\pr$.  The nonzero eigenvalues of $\bm Z^\prime \bm Z$ are the same as those $\bm Z\bm Z^\prime$, which when multiplied by $N T$,  equal the nonzero eigenvalues of $\bm X'\bm X$ and $\bm X\bm X^\prime$.

 We are interested in the low rank component of $\bm X$ viewed from the perspective of a factor model.
The static factor representation of the data is
    \begin{eqnarray}
    \label{eq:dgp}
    \bm X&=&\bm F\bm \Lambda^\prime  + \bm e.
    \end{eqnarray}
The common component $\bm C=\bm F\bm \Lambda\pr$ has reduced rank $r$ because $\bm F$ and $\bm \Lambda$ both have rank $r$.  Let $e_i^\prime=(e_{i1},e_{i2},...,e_{iT})$ and $e_t^\prime=(e_{1t},e_{2t},...,e_{Nt})$. The factor representation for
data of each unit $i$ is
\begin{eqnarray*}
 X_i&=&\bm F\Lambda_i+e_i.
\end{eqnarray*}
The $N\times N$ covariance matrix of $\bm X$ takes the form
\[  \bm\Sigma_X=\bm \Lambda\bm\Sigma_F \bm\Lambda^T + \bm\Sigma_e=\bm \Sigma_C+\bm \Sigma_e .\]
A {\it strict factor model} obtains when $\bm\Sigma_e$ is a diagonal matrix, which holds when the errors are cross-sectionally and serially uncorrelated. The classical factor model studied in \citet{anderson-rubin} uses the stronger assumption  that $e_{it}$ is iid and normally distributed. For economic analysis, this error structure is overly restrictive.  We work with the  {\it approximate factor model} formulated in \citet{chamberlain-rothschild}, which allows the idiosyncratic errors to  be weakly correlated in both the cross-section and time series dimensions. In such a case,  $\bm \Sigma_e$ need not be a diagonal matrix.

The defining characteristic of an approximate factor model is that the $r$ population eigenvalues of   $\bm \Sigma_C$  diverge with $N$ while all eigenvalues of $\bm \Sigma_e$ are bounded. Since $r$ can be consistently estimated, we will assume that $r$ is known. To simplify notation, the subscripts indicating that $\bm F$ is $T\times r$ and $\bm \Lambda$ is $N\times r$ will be suppresed when the context is clear.  Estimation of $\bm F$ and $\bm \Lambda$ in an approximate factor model with $r$ factors proceeds by minimizing  the sum of squared residuals:
 \begin{eqnarray*} \min_{\bm F,\bm \Lambda} \textsc{ssr}(\bm F,\bm \Lambda;r)&=&\min_{\bm F,\bm\Lambda} \frac{1}{NT} \|\bm X-\bm F\bm \Lambda^\prime\|_F^2\\ &=&\min_{\bm F,\bm\Lambda}\frac{1}{NT}\sum_{i=1}^N \sum_{t=1}^T (x_{it}-\Lambda_i^\prime F_t)^2.\end{eqnarray*}
 As  $\bm F$ and $\bm \Lambda$ are not separately identified, we impose  the normalization restrictions
\begin{equation}
 \frac{\bm F'\bm F}T=I_r, \quad  \quad  \frac{\bm \Lambda'\bm\Lambda}N \quad \text{is diagonal} .
\label{eq:normalization1}
\end{equation}
Even with these restrictions, the  problem is not convex and  is difficult to solve.
But we can   iteratively solve two bi-convex problems: (i) conditional on $\bm F$, minimizing the objective function with respect to  $\bm \Lambda$ suggests that  time series regressions of $X_i$ on $\bm F$ will give estimates of $\Lambda_i$ for each $i=1,\ldots N$; (ii) conditional on $\bm \Lambda$, doing $T$  cross-section regressions of $X_t$ on $\bm\Lambda$ will given estimates of $F_t$ for each $t$.  That is, we iteratively compute
\begin{subequations}
\begin{eqnarray}
 \tF &=& \bm X\tL(\tL'\tL)^{-1},\\
 \tLp &=& (\tF'\tF)^{-1}\tF' \bm X =\frac 1 T \tF' \bm X.
\end{eqnarray}
The solution upon convergence is the (static) asymptotic principal components (APC):
\begin{equation}
  (\tilde{\bm  F},\tilde{\bm \Lambda})=(\sqrt{T} \bm U_{NT,r},\sqrt{N}\bm V_{NT,r}\bm D_{NT,r}).
\end{equation}
\end{subequations}
Evidently, the solution involves eigenvectors because the algoirthm is an implementation of
 'orthogonal subspace iteration' algorithm for computing  eigenvectors, \citet[Algorithm 8.2]{golub-vanloan-3}. A related method is the 'alternating least squares' developed in \citet{deleeuw:04} and refined in \citet{unkel-trendafilov-10}  that treats  $\bm e$  as unknowns to be recovered.  Provided that  a low rank structure exists, the error bounds for these algorithms can be shown without  probabilistic assumptions about $\bm F,\bm \Lambda$, and $ \bm e$. We will need these assumptions to obtain distribution theory, and will treat  $\bm e$  as residuals rather than choice variables.

Analysis of the APC estimates in a setting of large $N$ and large $T$ must overcome two new challenges not present in the classical factor analysis of \citet{anderson-rubin}. The first  pertains to the fact that the errors are now allowed to be cross-sectionally correlated. The second pertains to the fact  that covariance matrix of $\bm X$ or $\bm X'$ are of dimensions $T\times T$ and $N\times N$ respectively, which are of infinite dimensions when $N$ and $T$ are large.   The asymptotic properties of the factor estimates were first studied in \citet{stock-watson-jasa:02,baing-ecta:02,bai-ecta-03}. Though the theory is well developed, the derivations are  quite involved.

In what follows, we will  establish the properties of  $\tilde F$ and $\tilde \Lambda$  using  simpler proofs and  under weaker assumptions than  previously used.  Throughout, we let
\[ \delta_{NT}=\min(\sqrt{N},\sqrt{T}).\]
Unless otherwise stated, $\|\bm A\|^2$ is understood to be the squared Frobenius norm of a $m\times n$ matrix $\bm A$.  That is, $\|\bm A\|^2=\|A\|_F^2=\sum_{i=1}^m\sum_{j=1}^n  |A_{ij}|^2=\text{Tr}(\bm A\bm A')$. The factor model can also be represented as
\[ X_{it}=\Lambda_i^\prime F_t+e_{it}.\]
A strict factor model assumes that $\mathbb E [e_{jt}e_{js}]=0$ for $s\ne t$. An approximate factor model relaxes this requirement.
\paragraph{Assumption A1:}  Let $\bm F^0$ and $\bm \Lambda^0$ be the true values of $\bm F$ and $\bm \Lambda$. Let $M<\infty$, not depending on $N$ and $T$.
     \begin{itemize}
    \item[i]  Mean independence: $ E(e_{it}|\lio ,\fto)=0$.

\item[ii] Weak  (cross-sectional and serial) correlation in the errors.
\begin{itemize}
\item[(a)] $E  \Big[\frac 1 {\sqrt{N}} \sumiN [e_{it}e_{is}-E(e_{it}e_{is})]\Big]^2 \le  M $,
\item[(b)] For all $i$ $,  \frac 1 T \sumtT\sumsT  | E (e_{it}e_{is})|  \le  M $,
\item[(c)] For all $t$, $\frac{1}{N\sqrt{T}} \|e_t^\prime \bm e\|=\OpdNTione$ and for all $i$, $\frac{1}{T\sqrt{N}} \|e_i^\prime \bm e\|=\OpdNTione$.
\end{itemize}
  \end{itemize}

  \paragraph{Assumption A2:}   (i) $\lim_{T\rightarrow\infty} \frac{\Fop \Fo}{T}=\bm \Sigma_F>0$; (ii); $\lim_{N\rightarrow\infty}\frac{\Lop\Lo}{N}=\bm \Sigma_\Lambda>0$; (iii) the eigenvalues of $\bm\Sigma_\Lambda \bm \Sigma_F$ are distinct.

\paragraph{Assumption A3:} (i) For each $t$, $E\|N^{-1/2}\sum_i \Lambda^0_i e_{it}\|^2\le M$ and $\frac{1}{NT} e_t^\prime\bm e^\prime\Fo=\OpdNTitwo$; (ii) for each $i$, $E\|T^{-1/2} \sum_t F^0_t e_{it}||^2\le M$  and
$\frac{1}{NT}e_i'\bm e\Lo=\OpdNTitwo$.

 Assumption A1 assumes  mean independence and some moment conditions.  Assumption A2  implies that $\|\Fo\|^2/T=O_p(1)$ and $\|\Lo\|^2/N=O_p(1)$, and that all $r$ eigenvalues of $\Lop\Lop$ diverge at the same rate of $N$. The   conditions ensure  a strong factor structure   which is  needed for identification.
Under Assumption A3,  the following holds:
\begin{eqnarray} \label{eq:FeeF}
\frac{1}{T}  \frac {\Fop\bm e\bm e'\Fo}{NT}&=& \frac 1 T \frac 1 N   \sum_{i=1}^N \Big[\bigg(\frac 1 {\sqrt{T}} \sum_t F^0_t e_{it}\bigg)\bigg(\frac 1 {\sqrt{T}} \sum_t F^0_t e_{it}\bigg)'\Big]
=O_p(1/T)\\
\label{eq:LeeL}
\frac{1}{N} \frac{\Lop\bm e'\bm e\Lo}{NT}&=&\frac{1}{N}\frac{1}{T} \sum_{t=1}^T \bigg[\bigg(\frac{1}{\sqrt{N}} \sum_i \Lambda^0_i e_{it}\bigg)\bigg(\frac{1}{\sqrt{N}} \sum_i\Lambda_i^0e_{it}\bigg)'\bigg]=O_p(1/N).
\end{eqnarray}

\begin{lemma}
Under Assumption A,
\label{lem:lemma-ee}
\begin{eqnarray*}
 \|\frac{\bm e\bm e'}{NT}\|^2
&=& O_p(\frac{1}{T})+O_p(\frac{1}{N})=O_p(\delta_{NT}^{-2}).
\end{eqnarray*}
\end{lemma}
Lemma \ref{lem:lemma-ee} establishes that  the normalized sum of squared covariances of the errors is of stochastic order that depends on the size of the panel in both dimensions. The proof comes from  observing that  $\bm e\bm e'$ is a $T\times T$ matrix  with $\sum_{j=1}^N e_{jt}e_{js}$ as its $(t,s)$ entry. Thus
\begin{eqnarray*}
\|\frac{\bm e\bm e'}{NT}\|^2&=& \frac{1}{N^2 T^2}\sum_{t=1}^T \sum_{s=1}^T \bigg(\sum_{j=1}^N e_{jt}e_{js}\bigg)^2 \\ &=&\frac{1}{T}\bigg[ \underbrace{\frac{1}{T}\sum_{t=1}^T \bigg(\frac{1}{N}\sum_{j=1}^N e_{jt}^2\bigg)^2}_{t=s}\bigg]+\frac{1}{N} \bigg[\underbrace{\frac{1}{T^2}\sum_{t=1}^T \sum_{s\ne t}^T \bigg(\frac{1}{\sqrt{N}}\sum_{j=1}^N e_{jt} e_{js}\bigg)^2}_{t\ne s}\bigg].
\end{eqnarray*}
 The first term is $O_p(1/T)$. The second term is $O_p(1/N)$ in the special case that $e_{jt}$ are serially uncorrelated. In general, the second term is $O_p(1/N)+O_p(1/T)$, which can be proved by adding and subtracting $E(e_{jt}e_{js})$ and use Assumption A1(ii)(b). Hence under Assumption A,  the idiosyncratic errors can only have limited time and cross-section correlations.

\section{Consistency Results}
From $\frac{1}{NT} \bm X\bm X^\prime=\bm U_{NT}\bm D_{NT}^2\bm U_{NT}^\prime$,
we have $ \frac{1}{NT}  \bm X\bm X' \tF =\tF \bm D_{NT,r}^2$.  Plugging in $\bm X=\bm F^0\bm \Lambda^{0\prime}+\bm e$ and expanding terms give
\begin{eqnarray}
\frac{\bm F^0(\bm\Lambda^{0\prime}\bm \Lambda^0)}{N}\frac{\bm F^{0'}\tF }{T}+\frac{\bm F^0\bm \Lop \bm e^{\prime} \tF}{NT} +
\frac{\bm e\bm \Lambda^0\bm  F^{0^\prime}\tF}{NT}
+ \frac{\bm e \bm e^{\prime}\tF}{NT} &=&\tF \bm D_{NT,r}^2 .
\label{identity1}
\end{eqnarray}
Various results will be obtained from this useful identity. Define the rotation matrix
\[\bm H_{NT,0}=\bigg(\frac{\bm \Lambda^{0^\prime}\bm \Lambda^0}{N}\bigg)\bigg(\frac{\bm F^{0^\prime}\tF}{T}\bigg) \bm D_{NT,r}^{-2}.\]
Note that this is the transpose of the one defined in \citet{baing-ecta:02}.

\subsection{Consistent Estimation of the Factor Space}
We want to establish that $\tilde F_t$ is close to $F_t$ and $\tilde\Lambda_i$ is close to $\Lambda_i$ in some well-defined sense.  Multiplying $\bm D_{NT,r}^{-2}$ to both sides of (\ref{identity1}) and using the definition of $\bm H_{NT,0}$, we have
\begin{eqnarray}
   \tF -\Fo \bm H_{NT,0}&=&\Big(\frac{\Fo\Lop\bm e^{\prime} \tF}{NT} +\frac{\bm e \Lo \Fop\tF}{NT}+ \frac{\bm e \bm e^{\prime}\tF}{NT} \Big) \bm D_{NT,r}^{-2}.
\label{eq:factorspace}
\end{eqnarray}
 Taking the norm on both sides.  we have
\begin{eqnarray*}
\frac 1 T  \|\tF -\Fo\bm H_{NT,0}\|^2 &\le& \left \{ 2 \Big(\frac {\|\Fo\|^2 \|\tF\|^2}{T^2} \Big) \Big(  \frac 1 T  \|\frac 1 N \Lop\bm e'\|^2 \Big)  + \frac {\|\tF\|^2} T \big\|\frac {\bm e \bm e^\prime} {NT} \big\|^2 \right \} \|\bm D_{NT,r}^{-2}\|^2,
\end{eqnarray*}

\begin{proposition} Under Assumption A, the following holds in squared
 Frobenius norm
\begin{eqnarray*}
(i). \quad\quad &&\frac{1}{T} \|\tF -\Fo\bm H_{NT,0}\|^2=\frac{1}{T} \sum_{t=1}^T\|\tilde F_t-\bm H_{NT,0}^{\prime}F_t^0\|^2
 =\OpdNTitwo\\
(ii). \quad\quad &&\frac{1}{N}\|\tL-\Lo (\bm H_{NT,0}')^{-1}\|^2=\frac{1}{N} \sumiN \|\tilde\Lambda_i- \bm H_{NT,0}^{-1}\Lambda_i^0\|^2=\OpdNTitwo\\
(iii). \quad\quad &&\frac{1}{NT} \|\tilde{\bm C}-\bm C^0\|^2=\frac{1}{NT}\sum_{i=1}^N\sum_{t=1}^T \|\tilde C_{it}-C^0_{it}\|^2=\OpdNTitwo.
\end{eqnarray*}

  \label{prop:prop1}
\end{proposition}
Part (i) of Proposition \ref{prop:prop1}   says  that the average squared deviation between $\tF$ and the space spanned by the true factors will vanish at rate $\min(N,T)$, which is the smaller of the sample size in the two dimensions. This result corresponds to Theorem 1 of \citet{baing-ecta:02}, but  the argument is now simpler. It  uses the fact that
   $\|\Fo\|^2/T=O_p(1)$ by Assumption A2, $\|\tF\|^2/T=r$ by normalization,  $\|\bm D_r^2\|=O_p(1)$, $ \frac 1 T  \|\frac 1 N \Lop\bm e'\|^2 = O_p(\frac 1 N )$ from equation (\ref{eq:LeeL})
and $ \|\frac 1 {NT}\bm e\bm e'\|^2 =O_p(\frac 1 T)+O_p(\frac 1 N)$ by Lemma \ref{lem:lemma-ee}.  Part (ii) follows by symmetry. Part (iii) does not depend on $\bm H_{NT,0}$ and is a consequence of (i) and (ii).

 Part (i)  is  weaker than uniform convergence of $\tilde F_t$ to $F_t^0$. However,  this result is sufficient to validate many uses of $\tilde F_t$, the most important being consistent estimation of the number of factors, and being able to treat $\tF$ as $\Fo$ in factor augmented regressions.

\subsection{The Limit of $\tFp\Fo/T$}
An important quantity in determining the properties of the factor estimates is $\tFp\Fo/T$.

\begin{proposition}
\label{prop:propQ}

 Let the $r\times r$ matrix $\bm \Sigma$ denote $\bm \Sigma =\bm \Sigma_{\Lambda}^{1/2}\bm \Sigma_F\bm \Sigma_{\bm \Lambda}^{1/2}$ and its spectral decomposition $\bm \Sigma = \bm \Upsilon \bm D^2_{r} \bm \Upsilon^\prime$ with $\bm \Upsilon' \bm \Upsilon=I_r$.
Under Assumption A,  then $\lim_{N,T\rightarrow\infty}\bm D^2_{NT,r}=\bm D^2_{r}$ and
 \[\tF'\Fop/T \pconv \bm Q=\bm D_{r} \bm \Upsilon \bm \Sigma_{\Lambda}^{-1/2}.\]
\end{proposition}

\paragraph{Proof.} The proof of $\lim_{N,T\rightarrow\infty}\bm D^2_{NT,r}=\bm D^2_{r}$ is given in \cite{stock-watson-di-wp}.  We focus on the limit of $\tF'\Fop/T$.  Multiply $\frac 1 T \bm F^{0\prime}$ on both sides of (\ref{identity1}), we have\footnote{Proposition \ref{prop:propQ} corresponds to Proposition 1 of \citet{bai-ecta-03} which is stated in terms of $\bm V$ instead of   $\bm D_r^2$. }
{\small
\begin{eqnarray*}
&&\bigg(\frac {\bm F^{0'}\bm F^0} T\bigg) \bigg(\frac{\bm \Lambda^{0'}\bm \Lambda^0} N \bigg)\bigg (\frac{\bm F^{0'}\tF } T\bigg)+\bigg(\frac{\bm F^{0'}\bm F^0} T\bigg) \bigg(\frac {\bm \Lambda^{0\prime}\bm e^{\prime}  \tF } {NT}\bigg)
  + \bigg(\frac {\bm F^{0'}\bm  e \bm \Lambda^0} {NT}\bigg)\bigg(\frac{ \bm F^{0'}\tF} T  \bigg)+ \bigg(\frac{ \bm F^{0'}e e^{\prime}\tF} {NT^2}\bigg)
\\ &&=\frac {\bm F^{0'}\tF} T \bm  D_{NT,r}^2.
\end{eqnarray*}
}
The second and  third terms on the left hand side  are negligible since  the $r\times r$ matrix
 \[  \frac {\bm F^{0'} \bm e \bm \Lambda^0} {NT} =\frac 1 {NT} \sum_i\sum_t F_t\Lambda_i' e_{it} =O_p(\delta_{NT}^{-2}) . \]
The fourth term is also negligible because
$ \frac { \Fop\bm e\bm e'\tF}{NT^2}=\frac {\Fop\bm e\bm e' \Fo}{NT^2} \bm H_{NT,0} + \frac {\Fop\bm e\bm e'(\tF-\Fo \bm H)}{NT^2} $ and each term is negligible. This implies that
\[ \bigg(\frac {\Fop\Fop} T\bigg) \bigg(\frac{\Lop\Lo} N \bigg)\bigg(\frac{\Fop\tF } T\bigg)+ o_p(1) =\frac {\Fop\tF} T  \bm  D_{NT,r}^2 \]
If we left multiply $(\bm \Lambda^{0'}\bm \Lambda^0/N)^{1/2}$ on each side and define
\begin{eqnarray*}
\bm \Sigma_{NT}&=&\Big(\frac{\Lop\bm \Lo} N \Big)^{1/2} \bigg(\frac {\Fop\Fo} T\bigg) \Big(\frac{\Lop\Lo} N \Big)^{1/2}, \\
\bar{\bm\Upsilon}_{NT}&=& \Big(\frac{\Lop\Lo} N \Big)^{1/2}\bigg(\frac{\Fop\tF } T\bigg),
\end{eqnarray*}
we have
\[ \bm \Sigma _{NT}  \bar{\bm\Upsilon}_{NT} +o_p(1) = \bar{\bm\Upsilon}_{NT}\bm D_{NT,r}^{2}. \]
Now  $\bar{\bm\Upsilon}_{NT}$ can be interpreted as the (non-normalized) eigenvectors of matrix $\bm\Sigma _{NT}$. These eigenvectors do not have unit length even asymptotically because $ \bar {\bm\Upsilon}^\prime_{NT}\bar{\bm\Upsilon}_{NT}\pconv  \bm D_r^2$. We can define  normalized eigenvectors $\bm\Upsilon_{NT}$ as  $\bm\Upsilon_{NT}=\bar{\bm\Upsilon}_{NT}\bm D^{-1}_{NT,r}$ so that $\bm\Upsilon_{NT}'\bm\Upsilon_{NT}\pconv I_r$. Since $\Lop\Lo/N\pconv \bm \Sigma_\Lambda$ and $\Fop \Fo/T\pconv \bm \Sigma_F$,  $\bm \Sigma_{NT}$ converges to $\bm \Sigma=\bm \Sigma_\Lambda^{1/2}\bm \Sigma_F\bm \Sigma_\Lambda^{1/2}$.  From $\bm \Sigma _{NT}  \bm\Upsilon_{NT} +o_p(1) = \bm\Upsilon_{NT}\bm D_{NT,r}^{2}$, taking the limit yields
$\bm \Sigma \bm \Upsilon =\bm \Upsilon \bm D_r^2$, where $\bm \Upsilon$ is the limit of  $\bm \Upsilon_{NT}$ (note that
since the eigenvalues of $\bm\Sigma$ are distinct,  $\bm\Upsilon$ is unique up to a column sign change, depending the column sign of
$\tilde {\bm F}$).  So $\bm D_r^2$ is the diagonal matrix consisting of the eigenvalues of $\bm \Sigma$, and  $\bm\Upsilon$ is the matrix of eigenvectors with $\bm \Upsilon ' \bm \Upsilon =I_r$.  We have
\[ \frac{\bm F^{0^\prime} \tF}{T} = \bigg(\frac{ \Lop\Lo}{N}\bigg)^{-1/2} \bm\Upsilon_{NT}\bm  D_{NT,r}
\pconv\bm \Sigma_\Lambda^{-1/2} \bm\Upsilon \bm D_{r}\equiv \bm Q^\prime.\]
Note that $\bm Q$ is not, in general, an identity matrix. Proposition \ref{prop:propQ} implies two useful results for what is to follow:
\begin{subequations}
\begin{eqnarray}
\bm Q' \bm D_r^{-2}&=&\bm \Sigma_\Lambda^{-1}\bm Q^{-1}
\label{eq:Qident1}\\
\bm \Sigma_F^{-1} \bm Q'&=& \bm Q^{-1}.
\label{eq:Qident2}
\end{eqnarray}
\end{subequations}
The first identity  follows from the definition of $\bm Q$ that $\bm Q'\bm D_r^{-2} \bm Q =\bm\Sigma_\Lambda^{-1/2}\bm \Upsilon^\prime \bm D_r^\prime \bm D_r^{-2} \bm D_r \bm \Upsilon\bm \Sigma_\Lambda^{-1/2}=\bm\Sigma_\Lambda^{-1}$.
The second identity uses
$\bm Q\bm \Sigma_F^{-1} \bm Q'=\bm D_r \bm \Upsilon' [\bm \Sigma_\Lambda^{-1/2}\bm \Sigma_F^{-1} \bm \Sigma_\Lambda^{-1/2}]\bm \Upsilon \bm D_r
=\bm D_r \bm \Upsilon' \bm \Sigma^{-1} \bm \Upsilon \bm D_r$ which simplifies to $\bm D_r \bm D_r^{-2} \bm D_r =I_r$.
The two identities can  equivalently be stated as $\bm Q^\prime\bm D_r^{-2} \bm Q=\bm \Sigma_\Lambda^{-1}$ and  $\bm Q\bm \Sigma_F^{-1} \bm Q'= I_r$, respectively.

\subsection{Equivalent Rotation Matrices}
As seen above, $\tF$ is based on $\bm U_r$,  the left singular vectors  of $\bm X$ and thus  all linear transformations of $\bm U_r$ are also solutions.
The following Lemma will be useful in establishing that $\bm H_{NT,0}$ has  asymptotically equivalent representations.
\begin{lemma}
\label{lem:Fte2}
Under Assumption A,    $\frac{\tF 'e'e\tF}{NT^2}=O_p(\delta_{NT}^{-2})$.
\end{lemma}
\noindent {\bf Proof:} From (\ref{eq:FeeF}), $\frac{\Fop ee'\Fo}{NT^2}=O_p(1/T)$. Now adding and subtracting terms,
\begin{eqnarray*}
\frac{\tF^\prime ee'\tF}{NT^2}&=&\frac{(\tF-\Fo\bm H)'ee'(\tF-\Fo\bm H)}{NT^2}\\&&+ \frac{\bm H\Fop ee'(\tF-\Fo H)}{NT^2}+ \frac{(\tF-\Fo \bm H)'\bm e\bm e'\Fo \bm H}{NT^2}+
\frac{\bm H'\Fop \bm e\bm e'\Fo \bm H}{NT^2}\\
&=& a+b+c+d.
\end{eqnarray*}
\begin{eqnarray*}
\|a\|&\le & \frac{\|\tF-\Fo \bm H\|^2}{T}\frac{\|\bm e\bm e'\|}{NT}=O_p(\delta_{NT}^{-2})O_p(\delta_{NT}^{-1})\\
\|b\|&\le & \frac{\|\tF-\Fo \bm H\|}{\sqrt{T}}\frac{\|ee'\|}{NT} \frac{\|\Fo \|}{\sqrt{T}}\|\bm H\|=O_p(\delta_{NT}^{-1})O_p(\delta_{NT}^{-1}) O_p(1)=O_p(\delta_{NT}^{-2})\\
\|b\|&\equiv & \|c\|\\
\|d\|&\le&
\|H\|^2 \frac{\|\Fop \bm e\bm e'\Fo \|}{NT^2}=O_p(\delta_{NT}^{-2}).
\end{eqnarray*}

We are now in a position to consider   asymptotically equivalent rotation matrices:

\begin{lemma}
 Let $\bm H_{NT,0}=\bigg(\frac{\bm \Lambda^{0^\prime}\bm \Lambda^0}{N}\bigg)\bigg(\frac{\bm F^{0^\prime}\tF}{T}\bigg) \bm D_{NT,r}^{-2}$ and define
 \[
\begin{array}{lll}
\bm   H_{NT,1}&=(\Lp\bm \Lambda^0)(\tL\pr \bm \Lambda^0)^{-1}, \quad
   &\bm H_{NT,1}^{-1}= (\tL\pr \bm \Lambda^0)(\Lp\bm \Lambda^0)^{-1},   \\
   \bm H_{NT,2} &=  (\Fp \bm F^0)^{-1}(\Fp \tF), \quad
   &\bm H_{NT,2}^{-1}= (\Fp \tF)^{-1} (\Fp \bm F^0)\\
   \bm H_{NT,3} &= (\tF\pr \bm F^0)^{-1}(\tF\pr \tF)=(\tF\pr \bm  F^0/T)^{-1} \quad
   &\bm H_{NT,3}^{-1}=(\tF'\bm F^0/T) =(\tF'\tF)^{-1} (\tF' \bm F^0)  \\
\bm   H_{NT,4} &=(\Lp \tL) (\tL^\prime\tL)^{-1} = (\Lp\tL/N) \bm D_{NT,r}^{-2},\quad
   &\bm H_{NT,4}^{-1} =\bm D_{NT,r}^2 (\Lp \tL /N)^{-1}.
\end{array}
\] Under Assumption A, the following holds for  $\ell=1,2,3,4$
\begin{itemize}
\item[i]  $ \bm H_{NT,\ell}=\bm H_{NT,0} + O_p(\delta_{NT}^{-2})$ ;
\item[ii]  $\bm H_{NT,\ell}\pconv  \bm Q^{-1}$.
\end{itemize}
\label{lem:lemmaH}
\end{lemma}
\noindent {\bf Proof:}
Part (ii) follows from  Proposition \ref{prop:propQ} that $\tFp\Fo/T\pconv \bm Q$. It remains to show that all alternative rotation matrices are asymptotically equivalent.

We begin with $\ell=1,3$. Recall that  $\bm D_{NT,r}^2$ is the matrix of eigenvalues of $\frac{\bm X\bm X\pr}{NT}$ associated with the eigenvectors $\tF$. Using the normalization $\tF'\tF=T \bm I_r$, we have
$ \tF\pr(\frac{\bm X\bm X\pr}{NT})\tF  = T \bm D_{NT,r}^2 $.
Substituting   $\bm X=\bm F^0\Lp + e$ into the above, we have
\begin{equation}
\label{eq:useful1} \bm D_{NT,r}^2 =\bigg(\frac{\tF\pr \bm F^0}{T}\bigg)
\bigg(\frac{\Lp \bm \Lambda^0}{N}\bigg) \bigg(\frac{\Fp \tF}{T}\bigg)
+  \underbrace{\frac 1 T\bigg( \frac{\tF\pr \bm e\bm e\pr\tF}{NT}\bigg)}_{\OpdNTitwo}+\OpdNTitwo
\end{equation}
where the last $\OpdNTitwo$ term represents the cross product term, which is dominated. The second on the right hand side
is  $\OpdNTitwo$ by Lemma {\ref{lem:Fte2}.
Substituting
$ (\frac{\Fp \tF}{T})^{-1}(\frac{\Lp\bm \Lambda^0}{N})^{-1}(\frac{\tF\pr \bm F^0}{T})^{-1} + \OpdNTitwo $ for $\bm D_{NT,r}^{-2}$ into $ \bm H_{0,NT}$  gives
\[ \bm  H_{NT,0} = \bigg(\frac{\tF\pr\bm F^0}{T}\bigg)^{-1} +\OpdNTitwo
.\]
  Next,
left and right multiplying $\bm X=\bm F^0\Lp +\bm  e$ by $\tF\pr$  and $\bm \Lambda^0$ respectively, dividing  by $NT$, and using $\tL =\tF\pr \bm X/T$, we obtain
\[ \frac {\tL \pr \bm \Lambda^0} N =\bigg(\frac {\tF\pr \bm F^0} T\bigg)\bigg( \frac{\Lp \bm \Lambda^0} N\bigg)  + \OpdNTitwo. \]
Substituting
$ \Big(\frac {\tL\pr\bm \Lambda^0} N\Big)^{-1} =\Big(\frac{\Lp \bm \Lambda^0} N\Big)^{-1}\Big(\frac {\tF\pr \bm F^0} T \Big)^{-1} +\OpdNTitwo
$
into $ \bm H_{NT,1}=(\frac{\Lp\bm \Lambda^0}{N})(\frac{\tL\pr \bm \Lambda^0}{N})^{-1}$,
we obtain
\[ \bm   H_{NT,1}= \Big(\frac {\tF\pr\bm F^0} T \Big)^{-1} +\OpdNTitwo .\]
Thus $\bm H_{NT,0}$ and $ \bm H_{NT,1}$ have the same asymptotic expression.

Now consider the case of $\ell=2,4$. From $\bm H_{NT,1}=\bm H_{NT,0}+\OpdNTitwo$,  we have
\[ \bigg(\frac{\Lp\Lo}{N}\bigg)\bigg(\frac{\tL\pr\Lo}{N}\bigg)^{-1} =\bigg(\frac{\tF' \bm F^0}{T}\bigg)^{-1} + \OpdNTitwo.\] Taking transpose and inverse,
and substituting  into the original definition of $\bm H_{NT,0}$ yield
\[ \bm H_{NT,0} =\bigg(\frac{\Lp\tL}{N}\bigg) \bm D_{NT,r}^{-2}  + \OpdNTitwo. \]
This proves part (iv). Now multiply  $\bm X=\bm F^0 \Lp +e $ by $\Fp$ on the left and $\bm\Lambda^0$ on the right  and divide by $NT$, we obtain
\[ \frac{\Fp \bm X \tL}{NT} =\frac{\Fp \bm F^0}{T} \frac{\Lp \Lapca}{N}  + \frac{\Fp \bm e \Lapca}{NT}. \]
Now
$\bm X\Lapca =\bm X\Lapca (\Lapca'\Lapca)^{-1} (\Lapca'\Lapca)=\tF(\Lapca'\Lapca)=\tF \bm D_{NT,r}^2 N $.
Thus
$ (\frac{\Fp \tF}{T})D_{NT,r}^2  =(\frac{\Fp \bm F^0}{T}) (\frac{\Lp \Lapca}{N}) +O_p(\delta_{NT}^{-2})$, or
equivalently,
\[ \bigg(\frac{\Fp \bm F^0}{T}\bigg)^{-1}\bigg(\frac{\Fp \tF}{T}\bigg) = \bigg(\frac{\Lp \Lapca}{N}\bigg) \bm D_{NT,r}^{-2} +O_p(\delta_{NT}^{-2}). \]
But the left hand side is equal to $\bm H_{NT,0}+O_p(\delta_{NT}^{-2})$.

These alternative rotation matrices,    first used  in \citet{baing-joe:19}, help understand  what is meant by consistent estimation of the factor space.  For example, since $\bm H_{NT,2}$ is obtained by regressing $\bm F_0$ on $\tilde{\bm  F}$,  $\bm H^\prime_{NT,1}F^0_t$ is asymptotically the fit from projecting  $\tilde F_t $ on the space spanned by $\bm F^0$. Similarly,  $\bm  H_{NT,1}$ is obtained by regressing $\bm \Lambda_0$ on $\tL$. Hence $\bm H_{NT,1}^{-1} \Lambda_i^0$ is asymptotically the fit from projecting $\tilde{  \Lambda}_i$ on the space spanned by $ \Lambda_i ^0$.

\section{Distritbution Theory}

Consider again $ X_i=\bm F^0 \Lambda^0_i+e_i$.
As we do not observe $\Fo$ or $\Lo$, we need an inferential theory for  $\tilde F_t$, $\tilde\Lambda_i$, and $\tilde C_{it}=\tilde F_t \tilde\Lambda_i^\prime$. The following assumption will be used to derive the limiting distributions.
\paragraph{Assumption B.} As  $N, T\rightarrow \infty$, the following holds  for each $i$ and $t$:
\begin{eqnarray*}
   \frac{1}{\sqrt{N}} \sum_{i=1}^N \Lambda^0 _i e_{it}&\dconv& \mathcal N(0,\bm \Gamma_t)\\
\frac{1}{\sqrt{T}}\sum_{t=1}^TF^0_te_{it}&\dconv& \mathcal N(0,\bm \Phi_i).
\end{eqnarray*}
Theorems 1 and 2 of \citet{bai-ecta-03} establish the limiting distribution of $\tilde F_t$ and $\tilde \Lambda_i$ based on the rotation matrix $\bm H_{NT,0}$ as follows:
\begin{subequations}
\begin{eqnarray}
\label{eq:FH0}
\sqrt{N}(\tilde F_t-\bm H_{NT,0}'F^0_t)&\dconv& \mathcal N( \bm D_{r}^{-2}\bm Q\bm \Gamma_t\bm Q'\bm D_{r}^{-2})\\
\sqrt{T}(\tilde \Lambda_i-\bm H_{NT,0}^{-1}\Lambda_i^0)&\dconv& \mathcal N(0,\bm Q^{'-1} \bm \Phi_i \bm Q^{-1}).
\label{eq:LH0}
\end{eqnarray}
\end{subequations}

We will use  alternative rotation matrices to obtain  the limiting distributions. To proceed, we need the following, shown in the Appendix.
\begin{lemma} \label{lem:lemma3}
Suppose that Assumption A holds. We have, for $\ell=0,1,2,3,4$,
\begin{itemize}
\item[i] $  \frac 1 T \Fp (\tF- \Fop \bm H_{NT,\ell})   =\OpdNTitwo$
\item[ii]
$   \frac 1 N  \Lp(\tL-\Lo\bm H_{NT,\ell}^{\prime -1})   =\OpdNTitwo$.
\item[iii]  $  \frac 1 T (\tF- \Fo \bm H_{NT,\ell})' e_i   =\OpdNTitwo $
for each $i$,
\item[iv]  $  \frac 1 N e_t'(\tL-\bm \Lo \bm H_{NT,\ell}^{\prime -1})   =\OpdNTitwo $  for each  $t$.

\end{itemize}
\end{lemma}

\bigskip

To obtain the  limiting distribution of $\tilde\Lambda_i$,
we  multiply $\frac 1 T \tF'$ to both sides of
$\bm X =\Fp\Lop +\bm e$ to obtain
\begin{eqnarray*}
 \frac 1 T \tFp\bm X &=&(\tFp \Fo/T) \Lop + \tFp\bm e /T \\
 \tLp &=& \bm H_{NT,3}^{-1} \Lop + \tFp\bm e /T \\
   &=&\bm H_{NT,3}^{-1} \Lop + \bm H_{3,NT}' \Fop \bm e/T + (\tF- \Fo \bm H_{NT,3})' \bm e /T .
\end{eqnarray*}
This implies
$ \tilde \Lambda_i -\bm H_{NT,3}^{-1} \Lambda_i^0 = \bm H_{NT,3}' \frac 1 T \sum_{t=1}^T F_t^0 e_{it} + \OpdNTitwo$.  For the  distribution of $\tilde F_t$,
we multiply  $\tL (\tLp\tL)^{-1}$ to  both  sides of
$\bm X= \Fo\bm \Lop +\bm e$:
\begin{eqnarray*}
\bm X \tL (\tLp\tL)^{-1}  &=&\Fo \bm \Lambda^{0'}\tL (\tLp\tL)^{-1}  + \bm e\tL (\tLp\tL)^{-1} \\
 \tF &=& \Fo \bm H_{NT,4}  +  \bm e\tL (\tLp\tL)^{-1}  \\
&=& \Fo \bm H_{NT,4}  +  \bm e \Lo \bm H_{NT,4}^{\prime -1}(\tLp\tL)^{-1} + e(\tL-\Lo \bm H_{NT,4}^{\prime -1}) (\tLp\tL)^{-1}
\end{eqnarray*}
This  implies
$ \tilde F_t-\bm H_{NT,4}' F_t^0 =(\frac{\tLp\tL}{N})^{-1} \bm H_{NT,4}^{-1 } \frac 1 N \sum_{i=1}^N \Lambda_i^0 e_{it} + O_p(\delta_{NT}^{-2})$.
Putting the results together,
\begin{subequations}
\begin{eqnarray} \label{eq:L-H3}
 \sqrt{T} (\tilde \Lambda_i -\bm H_{3,NT}^{-1} \Lambda_i^0) &=& \bm H_{NT,3}'\frac 1 {\sqrt{T}} \sum_{t=1}^T F_t^0 e_{it} + \sqrt{T} O_p(\delta_{NT}^{-2})\\
\label{eq:F-H4}
 \sqrt{N} (\tilde F_t -\bm H_{NT,4}' F_t^0) &=&\bigg(\frac{\tL'\tL}N\bigg)^{-1} \bm H_{NT,4}^{-1} \frac 1{\sqrt{N}} \sum_{i=1}^N \Lambda_i^0 e_{it} + \sqrt{N} O_p(\delta_{NT}^{-2}).
\end{eqnarray}
\end{subequations}
Assumption B then implies that $(\tF,\tL)$  are asymptotically normal with asymptotic variances given in (\ref{eq:FH0}) and (\ref{eq:LH0}). But  from  $\bm H_{NT,3}'=\bm H_{NT,2}^{-1} (\Fp \Fo/T)^{-1}$  and using (\ref{eq:L-H3}), it also holds that
\begin{eqnarray*}
 \sqrt{T} (\tilde \Lambda_i -\bm H_{NT,3}^{-1} \Lambda_i^0) &=& \bm H_{NT,2}^{-1} \bigg(\frac{\Fp \Fo}{T}\bigg)^{-1}  \frac 1 {\sqrt{T}} \sum_{t=1}^T F_t^0 e_{it} + \sqrt{T} O_p(\delta_{NT}^{-2}).
\end{eqnarray*}
Now since
$ (\tLp\tL/N)^{-1} \bm H_{NT,4}^{-1 } = (\Lp \tL /N)^{-1} =H_1^{\prime} (\Lop\Lo/N)^{-1}+O_p(\delta_{NT}^{-2})$,  we also have
\begin{eqnarray*}
  \sqrt{N}(\tilde F_t -\bm H_{NT,4}' F_t^0) &=&\bm H_{NT,1}' \bigg(\frac {\Lop\Lo} N\bigg)^{-1}  \frac{ 1}{\sqrt{ N}} \sum_{i=1}^N \Lambda_i^0 e_{it} + \sqrt{N} O_p(\delta_{NT}^{-2}).
\end{eqnarray*}
 Define
\begin{eqnarray*}
 \xi^F_{it}&=&\bigg(\frac{\Fp \Fo}{T}\bigg)^{-1}  \frac 1 {\sqrt{T}} \sum_{t=1}^T F_t^0 e_{it}\dconv \mathcal N(0,\bm \Sigma_F \bm \Phi_i\bm \Sigma_F),\\
\xi^\Lambda_{it}&=&\bigg(\frac {\Lop\Lo} N\bigg)^{-1}  \frac{ 1}{\sqrt{ N}} \sum_{i=1}^N \Lambda_i^0 e_{it}
\dconv \mathcal (0,\bm\Sigma_\Lambda^{-1} \bm \Gamma_t \bm\Sigma_\Lambda^{-1}),
\end{eqnarray*}
A compact way to summarize the estimation error  is
\begin{subequations}
\begin{eqnarray}
\label{eq:F41}
  \sqrt{N}(\tilde F_t -\bm H_{NT,4}' F_t^0)
&=& \bm H_{NT,1}' \xi^\Lambda_{it}+o_p(1)\\
\sqrt{T}(\Lambda_i-\bm H_{NT,3}^{-1} \Lambda_i^0)&=& \bm H_{NT,2}^{-1}\xi_{it}^F+o_p(1).
\label{eq:L32}
\end{eqnarray}
\end{subequations}

\begin{proposition}
\label{prop:propFL}
Under Assumptions A and B and the normalization that $\bm F'\bm F/T=I_r$ and $\bm\Lambda'\bm\Lambda/N$ is diagonal, we have,  as $N,T\rightarrow\infty$,
\begin{eqnarray*}
\sqrt{N}(\tilde F_t-\bm H_{NT,4}'F^0_t)&\dconv& \mathcal N(0, \bm Q^{\prime -1} \bm \Sigma_\Lambda^{-1}\bm \Gamma_t \bm\Sigma_\Lambda^{-1} \bm Q^{-1})\\
\sqrt{T}(\tilde \Lambda_i-\bm H_{NT,3}^{-1}\Lambda_i^0)&\dconv& \mathcal N(0, \bm Q\bm \Sigma^{-1}_F \bm \Phi_i \bm \Sigma^{-1}_F \bm Q^{\prime}).
\end{eqnarray*}
\end{proposition}

Although the limiting covariance matrices are different from those given in (\ref{eq:FH0}) and (\ref{eq:LH0}),  they are mathematically identical because of the different ways to represent $\bm Q$, as shown in  (\ref{eq:Qident1}) and (\ref{eq:Qident2}). Regardless of the choice of the rotation matrix, the factor estimates are all asymptotically normal. However,  as long as  $\tilde F$ are used as regressors, there is only one way to construct the confidence intervals in augmented regressions as  all rotation matrices are asymptotically the same.

It would seem convenient  to assume that $\bm H_{NT}$ is an identity matrix in making inference. But from Proposition \ref{prop:propQ}, any of the $\bm H_{NT}$  considered  is $\bm I_r$ only if the true $(\bm F^0,\bm \Lambda^0)$ satisfy $\frac{1}{T} \bm F^{0' }\bm F^0$ and $\bm \Lambda^{0'}\bm \Lambda^0$ is a diagonal matrix, which are strong identification assumptions. As pointed out in \citet{baing-joe:13}, these assumptions will affect not just where we center the limiting distribution of the factor estimates, but also their asymptotic variances.\footnote{It is  possible to relax some of these diagonality restrictions  so long as they are replaced by a sufficient number of linear restrictions as in  \citet{bai-wang-14}.}  Hence these restrictions are not innocuous.

 While there are many ways to represent the sampling error of $\tilde F_t$ and $\tilde \Lambda_i$,  the properties of $\tilde C_{it}$ are invariant to the choice of $\bm H_{NT,\ell}$, so we can simply write $\bm H_{NT}$. By definition, $C^0_{it}=\Lambda_i^{0\prime} F^0_t $ and $\tilde C_{it}=\tilde\Lambda_i^\prime \tilde F_t^0 $. Thus
\begin{eqnarray*}
\tilde C_{it}-C_{it}^0&=& \Lambda_i^{0\prime}\bm H_{NT}^{'-1} (\tilde F_t-\bm  H_{NT}'F_t^0)' +(\tilde\Lambda_i-\bm H_{NT}^{-1}\Lambda_i^0)^\prime \tilde F_t\\
&=&\Lambda_i^{0\prime}\bm H_{NT}^{'-1} (\tilde F_t-\bm  H_{NT}'F_t^0)' +F_t^{0'} \bm H_{NT}(\tilde\Lambda_i-\bm H_{NT}^{-1}\Lambda_i^0)+O_p(\delta_{NT}^{-2}).
\end{eqnarray*}
Using the results for $\tilde F_t$ and $\tilde\Lambda_i$,
\begin{eqnarray*}
(\tilde C_{it}-C_{it}^0)&=&\frac{1}{\sqrt{N}} \Lambda_i^{0\prime}\bm H_{NT}^{'-1} \bm H_{NT}'\bigg(\frac{\Lop\Lo}{N}\bigg)^{-1}  \frac 1{\sqrt{N}} \sum_{i=1}^N \Lambda_i^0 e_{it} \\
&&+\frac{1}{\sqrt{T}} F_t^{0'}\bm H_{NT} \bm H_{NT}^{-1} \bigg(\frac{\Fp \Fo}{T}\bigg)^{-1}  \frac 1 {\sqrt{T}} \sum_{t=1}^T F_t^0 e_{it} + \OpdNTitwo\\
&=&\frac{1}{\sqrt{N}} \Lambda_i^{0\prime} \xi^\Lambda_{it}+\frac{1}{\sqrt{T}} F_t^{0\prime} \xi^F_{it}+o_p(1).
\end{eqnarray*}
Now $F_t^{0\prime}  \xi^F_{it}\dconv N(0, W^F_{it})$ and $\Lambda_i^{0\prime} \xi^\Lambda_{it}\dconv N(0, W^\Lambda_{it})$,
where $W^F_{it}=F_t^{0\prime} \bm \Sigma_F^{-1}\bm \Phi_i \bm \Sigma_F^{-1} F_t^0$,
and $W^\Lambda_{it}=\Lambda_i^{0\prime} \bm \Sigma_\Lambda^{-1} \bm \Gamma_t \bm \Sigma_\Lambda^{-1} \Lambda_i^0$.
 This leads to a the distribution theory for the estimated common components.
\begin{proposition}
\label{prop:prop2}
Under Assumptions A and B and the normalization that $\bm F'\bm F/T=I_r$ and $\bm\Lambda'\bm\Lambda/N$ is diagonal, we have,  as $N,T\rightarrow\infty$,
\begin{eqnarray*}
\frac{\tilde C_{it}-C^0_{it}}{\sqrt{\frac{1}{N} \tilde W^\Lambda_{NT,it}+\frac{1}{T}\tilde W^F_{NT,it} }}&\dconv& \mathcal N(0,1)
\end{eqnarray*}
where  $\tilde W_{NT,it}^\Lambda$ and $\tilde W_{NT, it}^F$ are consistent estimates of $ W_{it}^\Lambda$ and $ W^F_{it}$, respectively.
\end{proposition}

Proposition \ref{prop:prop2} characterizes the sampling uncertainty of $\tilde C_{it}$ for each $i=1,\ldots, N$ and $t=1,\ldots T$. This error is also asymptotically normal but the convergence rate  is unusual:- it is the smaller of the sample size in the two dimensions, being $\delta_{N,T}=\min(\sqrt{N},\sqrt{T})$. The sampling distribution allows confidence intervals to be constructed for each or a collection of $\tilde C_{it}$. Such an analysis is possible because of Assumptions A and B.

The results thus far are derived for the APC estimates  where the principal components taken to be $\bm U_r$, where we recall that these are the left eigenvectors of  $\bm Z=\frac{\bm X}{\sqrt{NT}}$.   But some textbooks such as \citet{hastie-tibs-friedman} define principal components as $\bm U_r\bm D_r$. Though the two definitions will yield principal components that are perfectly correlated,  they are based on different normalizations. As normalizing $\bm F$ to be unit length can be restrictive for some purposes, \citet{baing-joe:19} define the principal components estimator (PC) as
\begin{subequations}
	\begin{eqnarray}
	\Fpca&=\sqrt{T} \, \bm U_{NT,r} \Dpca^{1/2}\label{eq:F-pca}\\
	\Lpca&=\sqrt{N}\, \bm V_{NT,r}  \Dpca^{1/2}  \label{eq:L-pca}.
	\end{eqnarray}
\end{subequations}
 The PC estimates are related to APC estimates:
\[\Fpca = \tF \Dpca^{1/2}, \quad \quad \Lpca = \tL \Dpca^{-1/2}.  \]
The limiting distribution of the PC estimates follow immediately from those  for $(\tF,\tL)$,
Why consider  the PC estimates? Because  $ \frac{\Fpca'\Fpca} T =\frac{ \Lpca'\Lpca} N = \bm D_{NT,r}$, so the factor estimates are no longer unit length. This opens the possibility for constrained  estimation. For example,  nuclear-norm regularization yields
\begin{equation}
(\Frpca_z,\Lrpca_z)=\argmin_{F,\Lambda} \frac 1 2 \Big( \norm{\bm Z-\bm F\bm \Lambda\pr}_F^2+\gamma\norm
{\bm F}_F^2+\gamma \norm{\bm \Lambda}_F^2\Big).\label{eq:rapca-obj}
\end{equation}
This set up is of interest because it is a convexifed formulation of the {\em minimum-rank} problem which has a  long standing history in factor analysis and has received renewed interest in the machine learning literature in recent years. See  \citet{tenberge-kiers},  \citet{scpw} and \citet{bcm:17} among others. The solution entails truncating small singular values.
Define the singular value thresholding operator (SVT) as
\begin{equation}\bm D_{NT,r}^\gamma=\bigg[\bm D_{NT,r}-\gamma I_r\bigg]_+\equiv
\max(\bm D_{NT,r}-\gamma \bm I_r,0).
\label{eq:soft}
\end{equation}
The robust principal components estimator (RPC) is defined as:
\begin{subequations}
\begin{eqnarray}
\label{barF-X} \Frpca&=& \sqrt{T}\bm U_{NT,r} (\bm D^\gamma_{NT,r})^{1/2}\\
\label{barL-X}
\Lrpca &=&\sqrt{N}\bm V_{NT,r} (\bm D^\gamma_{NT,r})^{1/2}.
\end{eqnarray}
\end{subequations}
Since $(\Frpca,\Lrpca)=(\tilde{\bm F} (\bm D_{NT,r}^\gamma)^{1/2},\tilde{\bm \Lambda}\bm \Delta_{NT})$ where $\bm \Delta_{NT}^2=\bm D_{NT,r}^\gamma \bm D_{NT,r}^{-1}$. This penalized objective function can be used to obtain   a robust estimate of the number of factors.

\section{The Number of Factors}
The foregoing results presume that the number of factors $r$ is unknown which is not usually the case in practice.
An informal analysis  is to plot the eigenvalues and use the point where the plot changes slope  as an estimate of $r$. This is    the `scree plot' first considered in \citet{cattell:66} and implemented in many software packages. A more formal approach is to balance the cost of adding an additional factor against model complexity. Let    $\text{ssr}(\tF,k)$ be the sum of squared residuals when $k$   factors are estimated.  For given $r_{\max}$,  \citet{baing-ecta:02} propose to determine $r$ by
\[ \hat r=\min_{k=0,\ldots, \text{rmax}} IC(k), \quad\quad \widetilde{IC}(k)= \log(\text{ssr}(\tilde {\bm F},k)) + k \cdot g(N,T)\]
where  $g(N,T) $ is chosen such that
\[ (i). \quad g(N,T)\rightarrow 0, \quad \quad (ii). \quad \delta_{NT}^2 g(N,T)\rightarrow \infty.\]
The original proof of Lemma 3 in \citet{baing-ecta:02} is based on  $\bm H_{NT,0}$ matrix and is tedious. But from  $\tilde e= X-\tilde C$, it follows
from Assumptions A and B that  for any fixed $k\ge r$,
 \[\frac{1}{NT}\sum_{i=1}^N \sum_{t=1}^T \tilde e_{it}^2-\frac{1}{NT}\sum_{i=1}^N \sum_{t=1}^T e_{it}^2=\OpdNTitwo.
\]
This  implies that
\[ \frac{1}{NT} \bigg(\text{ssr}(\tF,k)-\text{ssr}(\Fo\bm H,r)\bigg)=O_p(\delta_{NT}^{-2}).\]
For $k<r$, \citet{baing-ecta:02} shows that, for some $c>0$,
\[ \frac{1}{NT} \bigg(\text{ssr}(\tF,k)-\text{ssr}(\Fo\bm H,r)\bigg) \ge c.\]
These results imply that  $g(N,T)=\frac{\log \delta^2_{NT}}{\delta^2_{NT}}$ is appropriate, as are
$(\frac{N+T}{NT}) \log \delta_{NT}^2$ and $(\frac{N+T}{NT}) \log ( \frac{NT}{N+T})$ since they satisfy the two conditions.

 To relate the criterion function above to eigenvalues,  recall that by construction, the standardized data have the property that  $\norm{\bm Z}_F^2=d_1^2+d_2^2 +\cdots +d_{\min\{N,T\}}^2=1$.  The PC estimate of a  low rank component $\widehat{\bm  C}_k$  assumed to be of rank $k$ satisfies
 \[||\widehat{\bm  C}_k||^2_F=\norm{\bm D_{NT,k}^2}=d_1^2+d_2^2+\cdots +d_k^2.\]
 Then \textsc{ssr}$_k$ based on PC estimates can be written as
\[\textsc{ssr}_k  =1-\sum_{j=1}^k d_j^2=\norm{\bm Z-\widehat {\bm C}_k}_F^2,\]
showing that  criteria  in the $IC$ class are also based on eigenvalues.  \citet{ahn-horenstein:13} consider  successive changes in eigenvalues while \citet{onatski-ecta:09} which formalizes the scree  plot of \citet{cattell:66}. It is difficult to avoid using eigenvalues to determine $r$.

Recall that   the number of strong factors in an approximate factor model is the number of eigenvalues that increase with $N$.  To take the focus on strong factors one step further,  \citet{baing-joe:19} use the
 rank-regularized PC estimates  $\norm{\bm{\overline C}_k}=\norm{\bm D_{NT,k}^\gamma}_F^2$ in the IC  criterion function.   Given $k$ and $\gamma>0$, the regularized sum of squared residuals is
$ \widehat{\textsc{ssr}}_k(\gamma)= 1-\sum_{j=1}^k (d_j-\gamma)_+^2=\norm{\bm Z-\overline {\bm C_k}}_F^2.$
This leads to a class of rank-regularized  class of  criteria
\begin{equation}
\bar r=\min_{k=0,\ldots,\text{rmax}}
\log\bigg(1-\sum_{j=1}^k (d_j-\gamma)_+^2\bigg) + k g(N,T) . \label{eq:IC-17}
\end{equation}
Taking the approximation $\log(1+x)\approx x$, we see that
\[ \overline{ IC}(k)=\widehat{IC}(k)+\gamma \sum_{j=1}^k\frac{(2d_j-\gamma)}{\widehat{\textsc{ssr}}_k}.
\]
Since $d_j\ge d_j-\gamma\ge 0$, the penalty is heavier in $\overline{IC}(k)$ than $\widehat {IC}(k)$. The rank constraint adds a data dependent term to each  factor to deliver  a more conservative estimate of $r$ that does not require the researcher to make precise  the source of the small singular values. They can be due to genuine weak factors, noise corruption, omitted lagged and non-linear interaction of the factors that are of lesser importance.



\subsection{Linear Constraints}
The minimization problem in (\ref{eq:rapca-obj}) has a unique solution under the normalization
 $\bm F\pr \bm F =\bm \Lambda\pr \bm \Lambda =\bm D_r $. However, the unique solution may or may not have economic interpretations.
This section considers
$m$ linear restrictions on $\bm \Lambda$ of the form
\begin{equation} \label{eq:restriction} \bm R \, \vec(\bm \Lambda) = \phi   \end{equation}
 where $\bm R$ is $m \times Nr$, and $\phi$ is $m\times 1$. Both $\bm R$ and $\phi$ are assumed known  a priori. Economic theory may suggest a lower triangular $\bm \Lambda$. By suitable design of $\bm R$,  causality restrictions can be expressed as $\bm R\, \text{vec}(\bm \Lambda)=\phi$ without ordering the data a priori.  Cross-equation restrictions  such as due to  homogeneity of the loadings across individuals or a subgroup of individuals suggested by theory can also be considered. Other restrictions are considered in \citet{stock-watson:handbook-16}.  Nos imposing  diagonality of $\bm F\pr \bm F$ and $\bm \Lambda\pr\bm \Lambda$ for identification  (rather than statistical normalizations) actually generate  linear  constraints on the loadings (\ref{eq:restriction}) that can be used as  over-identifying restrictions with which we can use to test economic hypothesis.
 The Appendix provides an example how to implement the restrictions in \textsc{matlab}.

The linear restrictions on the loadings we consider here are known a priori. This stands in contrast to  sparse principal components  (SPC) estimation that either imposes  \textsc{lasso} type penalty on the loadings, or shrinks the individual entries to zero in a data dependent way.\footnote{For SPC, see  \citet{jolliffee-trendafilov-uddin}, \citet{ma:13}, \citet{shen-huang:08}, and \citet{zht}. The SPC is in turn different from the POET estimator of \citet{fan-liao-mincheva}  which constructs the principal components from a matrix that shrinks the small singular values towards zero.}

The constrained  factor estimates $(\Frpca_{\gamma,\tau},\Lrpca_{\gamma,\tau})$ are defined as solutions to the penalized problem
\begin{equation}
\label{eq:constrained-obj}
(\Frpca_{\gamma,\tau},\Lrpca_{\gamma,\tau})=\min_{F,\Lambda} \frac{1}{2} \norm
{\bm Z-\bm F\bm \Lambda'}_F^2+ \frac{\gamma }{2}\bigg(\norm{\bm F}_F^2+\norm
{\bm \Lambda}_F^2\bigg)+\frac{\tau}{2} \norm{\bm R \, \vec(\bm \Lambda) -\phi}_2^2
\end{equation}
where $\gamma$ and $\tau$ are regularization parameters.    The linear constraints can be imposed with or without the rank constraints. Imposing cross-equation restrictions will generally require iteration till the constraints are satisfied.

 The first order condition with respect to $\bm F$ for a given $\bm \Lambda$ is unaffected by the introduction of the linear constraints on $\bm \Lambda$. Hence, the solution
\begin{equation}
\label{eq:Fc-given-L}
\bar {\bm F}_{\gamma,\tau} = \bm Z\bm  \Lambda(\bm \Lambda\pr \bm \Lambda+\gamma \bm I_r)^{-1},\quad \forall\tau \ge 0
\end{equation}
 can be obtained from a ridge regression of $\bm Z$ of $\bm \Lambda$.
\
To derive the first order condition with respect to $\bm \Lambda$, we rewrite the problem in vectorized form:
\[ \|\bm Z-\bm F\bm \Lambda'\|^2_F=\|\vec(\bm Z')-(\bm F\otimes \bm I_N)\vec(\bm \Lambda)\|_2^2, \quad \|\bm \Lambda\|_F^2 =\|\vec(\bm \Lambda)\|_2^2. \]
The first order condition with respect to $\vec(\Lambda)$ is
\begin{eqnarray*}
0&=& -(\bm F'\otimes \bm I_N)\Big[\vec(\bm Z')-(\bm F\otimes \bm I_N)\vec(\bm \Lambda)\Big]+ \gamma \, \vec(\bm \Lambda) + \tau \bm R'[\bm R \, \vec(\bm \Lambda)-\phi] \\
&=& -\vec(\bm Z'\bm F)-\tau \bm R'\phi +(\bm F'\bm F\otimes \bm I_N) \, \vec(\bm \Lambda) + \gamma  \,\vec(\bm \Lambda ) +\tau \bm R'\bm R \,\vec(\bm \Lambda).
\end{eqnarray*}
Solving for $\vec(\bm \Lambda)$ and  and denoting the solution by $\vec(\bar {\bm \Lambda}_{\gamma,\tau})$, we  obtain
\begin{eqnarray}  \label{additional_penalty}
\Lrpca_{\gamma ,\tau} &=&\Big((\bm F'\bm F\otimes \bm I_N) +\gamma \bm  I_{Nr} +\tau  \bm R'\bm R \Big)^{-1} \Big[ \vec(\bm Z'\bm F)+\tau \bm R'\phi\Big] \\
&=&\Big((\bm F'\bm F +\gamma \bm I_r)\otimes \bm I_N +\tau \bm R^\prime \bm R\Big)^{-1}\bigg[ \vec(\bm Z'\bm F+\tau \bm R^\prime \phi)\bigg] \nonumber
\end{eqnarray}
where the last line follows from the fact that $(\bm F'\bm F\otimes \bm I_N) +\gamma  \bm I_{Nr}=(\bm F'\bm F+\gamma  I_r) \otimes \bm I_N$.
Equations (\ref{eq:Fc-given-L}) and (\ref{additional_penalty}) completely characterize the solution under rank and linear restrictions. In general, the solution will need to be solved by iterating the two equations until convergence. A reasonable starting value is $(\Frpca,\Lrpca)$, the solution satisfying the rank constraint and before the linear restrictions are imposed. However, while $\Frpca\pr \Frpca=\Lrpca\pr \Lrpca=\bm D_r^\gamma$ and $\bm D_r^\gamma$ is diagonal,  $\bar{\bm  F}_{\gamma,\tau}^\prime \bar {\bm F}_{\gamma,\tau}$ and $\bar {\bm \Lambda}_{\gamma,\tau}^\prime \bar {\bm \Lambda}_{\gamma,\tau} $ will  not, in general, be diagonal when linear restrictions are present.

These constraint will not bind unless $\tau=\infty$, and  we denote by $\Lambda_{\gamma,\infty}$ the binding solution. Observe that  in the absence of linear constraints (i.e. $\tau=0$),
\begin{equation} \label{eq:ridge}
  \vec(\Lridge) =\Big((\bm F'\bm F +\gamma \bm I_r)\otimes \bm I_N \Big)^{-1} \vec(\bm Z'\bm F) \end{equation}
which is a ridge  estimator. Furthermore, (\ref{eq:Fc-given-L}) and (\ref{eq:ridge}) are the RPCA estimates when iterated till convergence.   An estimator  that satisfies  both the rank constraint and $\bm R \, \vec(\bm \Lambda)=\phi$ can be obtained as follows.
For given $\bm F$, let $\Lcpca$ be the solution to (\ref{eq:constrained-obj}) with $\tau=\infty$. Also let $\Lridge$ be the solution  with $\tau=0$.
Similar to the usual formula for restricted OLS, the restricted solution is related to the unrestricted one  as follows:
\begin{eqnarray}
\label{RR-simplified}
\vec(\bm \Lcpca)&=&\vec(\bm \Lridge)- \nonumber\\
&&[(\bm F'\bm F+\gamma  \bm I_r)^{-1}\otimes \bm I_N] \bm R'
 \cdot  \Big[\bm  R[(\bm F'\bm F+\gamma \bm  I_r)^{-1}\otimes \bm I_N] \bm R'\Big]^{-1}\Big( \bm R \, \vec(\bm \Lridge)-\phi\Big)   \nonumber\\
\end{eqnarray}
This implies that a restricted estimate of $\bm \Lambda$ that satisfies both the rank and linear restrictions can be obtained by imposing the linear restrictions on $\bar{\bm \Lambda}_{
\gamma,0}$, the RPCA solution of $\bm \Lambda$ that only imposes rank restrictions. It is easy to verify $\Lcpca$ satisfies  restriction (\ref{eq:restriction}). Once the restricted estimates are obtained, $\bm F$ needs to be re-estimated based on (\ref{eq:Fc-given-L}). The final solution is obtained by iterating
(\ref{eq:Fc-given-L}) and (\ref{RR-simplified}).  We note again that $\bar{\bm F}_{\gamma,\infty}^\prime \bar{\bm F}_{\gamma,\infty}$ and $\bar {\bm \Lambda}_{\gamma,\infty}^\prime \bar{\bm \Lambda}_{\gamma,\infty} $ will not, in general, be diagonal matrices in the presence of linear restrictions.

\section{Conclusion}
This note has presented simplified proofs for properties of the factor estimates by principal components under the assumption that the factors are strong ie.   $\bm \Lambda^\prime\bm \Lambda/N>0$ and the population eigenvalues of $\bm \Sigma_X$  increase with $N$.
Situations may arise that require a precise documentation of the number of factors, whether  they are strong or weak. \citet{onatski-joe:12} formalizes   weak factors as those with loadings satisfying $\bm \Lambda'\bm \Lambda>0$  as $N$ and $T$ tend to infinity, and  so the population eigenvalues of $\bm \Sigma_X$  increase slower than $N$. The model choice of strong versus weak factors depends on the objective and the assumptions that the researcher finds defensible.  We have also focused  exclusively on estimation of  static factors.  Dynamic principal components are analyzed in \citet{fhlr-restat,fhlr-joe04}. 


\newpage
\section*{Appendix}

\section*{Proof of Lemma \ref{lem:lemma3}}
\paragraph{Proof of (i).}
Let $\bm H_{NT}=\bm H_{4,NT}$. Then
\begin{eqnarray*}
\tF-\Fo\bm H_{NT}&=&\bm e\tL(\tLp\tL)^{-1}=\frac{\bm e\tL}{N}
\bigg(\frac{\tLp\tL}{N}\bigg)^{-1}=\frac{\bm e\tL}{N}\bm D_{NT,r}^{-2}.
\end{eqnarray*}
Hence $\frac{1}{T} \Fop(\tF-\Fo\bm H_{NT})=\frac{1}{NT} \Fop\bm e\tL \bm D_{NT,r}^{-2}=a+b$, where
\begin{eqnarray*}
a&=& \frac{1}{NT} \Fop\bm e\Lo \bm H_{NT}{'^{-1}}\bm D_{NT,r}^{-2}
=\bigg(\frac{1}{NT}\sum_{t=1}^T \sum_{i=1}^N F_t^0\Lambda_i^{0'} e_{it}\bigg)\bm H_{NT}^{'^{-1}} \bm D_{NT,r}^{-2}=O_p(\delta_{NT}^{-2})\\
\|b\|&=&\|\frac{1}{NT} \Fop\bm e(\tL-\Lo \bm H_{NT}^{'^{-1}})\|
\le  \frac{1}{\sqrt{N}} \|\tL-\Lo\bm H{_{NT}}^{'^{-1}}\| \frac{1}{\sqrt{N} T}\|\Fop\bm e\|\\
&= & O_p(\delta_{NT}^{-1})O_p(\frac{1}{\sqrt{T}})=\OpdNTitwo
\end{eqnarray*}
where we use $\frac 1 {\sqrt{N} T}\|\Fop\bm e\|=O_p(T^{-1/2})$ by equation (\ref{eq:FeeF}).
The proof of (ii) follows by symmetry to part (i).

\paragraph{Proof of (iv):}
Here, we use $\bm H_{NT}=\bm H_{NT,3}$. Then
\[ \tLp- \bm H_{3,NT}^{-1}\Lop = (\tFp\tF)^{-1}\tFp \bm X-(\tFp\tF)^{-1} \tFp \Fo \Lo=\tFp \bm e/T  \]
and $ \frac{1}{N}e_t^\prime(\tL-\Lo\Hpi)= \frac 1 {NT} e_t' \bm e' \tilde {\bm F} = a+b$, where
$a = \frac 1 {NT} e_t' \bm e' \bm F^0 \bm H_{NT}$ and $ b= \frac 1 {NT} e_t'\bm e'(\tF-\Fo\bm  H_{NT})$.
By Assumption A3,  $a=\OpdNTitwo$, and
\begin{eqnarray*}
\|b\|  &\le &\frac{1}{N\sqrt{T}}\|e_t^\prime \bm e\| \|\tF-\Fo\bm  H_{NT}\|/\sqrt{T}=\OpdNTione\OpdNTione=\OpdNTitwo,
\end{eqnarray*}
where $\frac{1}{N\sqrt{T}}\|e_t^\prime \bm e\|=\OpdNTione$ by Assumption A1(ii)(c).
Proof of (iii) follows by symmetry.

\newpage

\bibliographystyle{harvard}
\bibliography{factor,metrics,metrics2,bigdata,macro}
\end{document}